\documentclass[runningheads]{llncs}
\usepackage[T1]{fontenc}
\usepackage{graphicx}
\usepackage[noend]{algpseudocode}
\usepackage{algorithm,algorithmicx}
\usepackage{amsmath}
\usepackage{amssymb}
\usepackage{todonotes}
\usepackage{hyperref}
\usepackage{wrapfig}
\usepackage{array, multirow, caption, booktabs}
\newcolumntype{L}[1]{>{\raggedright\arraybackslash}p{#1}}
\usepackage{tikz}
\newcolumntype{Y}{>{\raggedright\arraybackslash}X} 
\usepackage{bm}
\newcommand{\bb}[1]{\bm{\mathrm{#1}}}

\begin{document}
\title{Self-supervised learning of inverse problem solvers in medical imaging}
%
%
\author{
Ortal Senouf*\inst{1} \and
Sanketh Vedula*\inst{1} \and
Tomer Weiss\inst{1}\and\\
Alex Bronstein \inst{1} \and
Oleg Michailovich \inst{2} \and
Michael Zibulevsky \inst{1}
}
\authorrunning{O. Senouf et al.}

\institute{Technion, Israel \and
University of Waterloo, Canada \\
* -- contributed equally}
\maketitle             
\begin{abstract}
In the past few years, deep learning-based
methods have demonstrated enormous success for solving inverse problems in medical imaging. In this work, we address the following question:
\textit{Given a set of measurements obtained from real imaging
experiments, what is the best way to use a learnable model and the physics of the modality
to solve the inverse problem and reconstruct the latent image?} Standard supervised learning based methods approach this problem by collecting data sets of known latent images and their corresponding measurements. However, these methods are often impractical due to the lack of availability of appropriately sized training sets, and, more generally, due to the inherent difficulty in measuring the "groundtruth" latent image. In light of this, we propose a self-supervised approach to training inverse models in medical imaging in the absence of aligned data. Our method only requiring access to the measurements and the forward model at training. We showcase its effectiveness on inverse problems arising in accelerated magnetic resonance imaging (MRI).

\end{abstract}
\section{Introduction}
In the past years, there has been an enormous success in deploying deep learning-based methods in imaging, image processing, and computer vision. Most of these tasks, if tackled as a supervised learning problem, require collecting a large dataset of measurements and their corresponding latent variables, which would be referred to as labels in classification and detection tasks, and as ground truth in regression tasks. 
Whereas the task of labeling images, albeit not simple, can be addressed by using a large number of human annotators, the task of collecting measurements and their corresponding aligned ground-truth images is much harder and often impractical. The acquisition of a groundtruth image typically requires subjecting the same object of interest to a different imaging modality or to the same modality configured to provide more accurate measurements. The need to register such images at the sub-pixel level, coping with the object's natural deformation is often very difficult to surmount.

The limitation of the supervised regime is the main motivation of the present paper. We focus on answering the following question: Given a \emph{single} measurement obtained from a real imaging system, and our knowledge about the forward model embodying the physics of the imaging modality, can we learn an operator solving the inverse problem and reconstructing the latent signal?
We henceforth refer to this learning regime as \emph{self-supervised learning} (SSL). It is important to emphasize that the proposed self-supervised learning is cardinally different from \emph{unsupervised learning}, despite the fact that no groundtruth is used in both cases. While the latter relates mostly to generative models which try to estimate the latent data distributions, SSL aims at solving the inverse problem by exploiting internal information within the measurements themselves, and more specifically in our case, trying to explain or dissect the given measurements.  

\textbf{Contributions.} We propose an SSL framework, comprising two building blocks: a convolutional neural network (CNN) that serves as the prior, and a forward model, embodying the imaging physics into the pipeline. Several recent studies \cite{ulyanov2018deep,shocher2018zero} demonstrated  that a CNN can serve as a good prior for a wide range of images classes -- a line of works that is generally referred to as \emph{deep image prior}. From this perspective, the present solution can be viewed as the embodiment of deep image priors in general inverse problems.  
We demonstrate and evaluation our method on the case of accelerated magnetic resonance imaging. The forward model in that case is the MR $k$-space sampling trajectory. By applying SSL to this task, we introduce a significant improvement (around $2-3$dB PSNR) compared to an off-the-shelf total variation-regularized solver, and even some level of proximity to the performance of the fully supervised restoration model.            

\section{Methods}
\label{sec:methods}
In this work, we are interested in \textit{inverse problems}, which aim to calculate, from the measurements, the latent signal that produced them. The process of measuring the latent signal is referred to as the \textit{forward model}. We denote the forward model with the operator $\mathcal{F}(\cdot)$ that maps the entities in the domain of latent signals to the measurements. Many types of inverse problems arising in signal and image processing and medical imaging involve a \textit{linear} forward model, which can be straightforwardly expressed as the matrix product
\begin{equation}
    \label{eq:inv_problems}
    \mathbf{y} = \mathbf{F}\mathbf{x} + \eta.
\end{equation}
Here $\mathbf{x}\in \mathbb{R}^n$ is the latent signal that is measured through the forward model $\mathbf{F} \in \mathbb{R}^{m \times n}$, resulting in the observed measurement $\mathbf{y}\in\mathbb{R}^m$; $\eta$ denotes additive measurement noise. An inverse problem consists of estimating the latent signal $\mathbf{x}$ from the measurement $\mathbf{y}$. 

Several important inverse problems admit the above structure, for example
\begin{itemize}
    \item \textit{Denoising}: $\mathbf{F}$ is an $n \times n$ identity matrix $\mathbf{I}$.
    \item \textit{Inpainting}: $\mathbf{F}$ is an identity matrix with missing entries.
    \item \textit{Compressed sensing:} $\mathbf{F}$ is an $m \times n$ ($m \ll n$) random matrix (typically with Gaussian or super-Gaussian i.i.d. entries).
    \item \textit{Tomography:} $\mathbf{F}$ is the Radon transform (line integral) matrix.
\end{itemize}
Linear modeling of $\mathcal{F}$ is not accurate for a range of more exotic problems arising in computational and medical imaging such as multiple-scattering computed tomography, optical diffraction tomography, and wave-propagation inverse problem in ultrasound imaging. The proposed methodology applies to these modalities as well as long as the appropriate forward operator $\mathcal{F}$ is known. Therefore, in order to emphasize the  broader applicability of the proposed framework, we will refer to the forward model as $\mathcal{F}$ instead of the matrix $\mathbf{F}$.

\begin{figure}[!htb]
    \centering
    \includegraphics[width=0.7\textwidth]{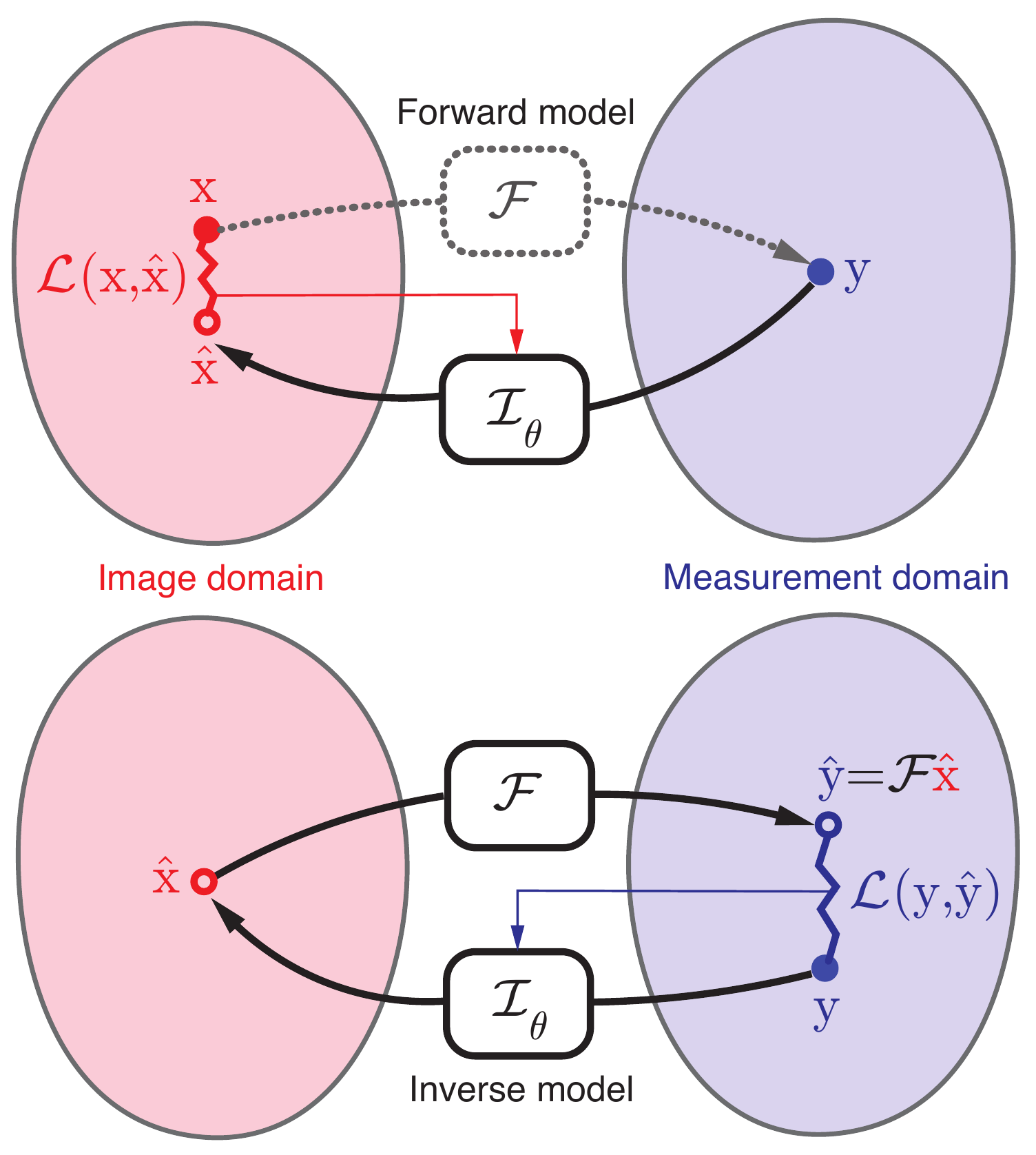}
    \caption{{\small Comparison of different approaches to learning inverse models. In the standard supervised approach (top), many pairs of latent images $\mathbf{x}$ and corresponding measurements $\mathbf{y}$ are available at training, and a loss in the image domain drives the parameters $\theta$ of the inverse model such that $\hat{\mathbf{x}} = \mathcal{I}_\theta(\mathbf{y})$ is close to $\mathbf{x}$. In the proposed self-supervised approach (bottom) only access to the measurements $\mathbf{y}$ and the forward model $\mathcal{F}$ is assumed. The loss is formulated in the measurement domain, and the inverse model is trained such that $\mathcal{F}(\mathcal{I}_\theta(\mathbf{y}))$ is close to $\mathbf{y}$. }}
    \label{fig:framework}
\end{figure}

\subsection{Prior-based solvers}
\label{sec:prior}
One of the standard formulations of inverse problems is in the form of maximum \textit{a posteriori} (MAP) estimation the latent signal $\mathbf{x}$ from the measurements $\mathbf{y}$. This formulation allows to introduce information about the latent image through the prior $P_X (\mathbf{x})$, and boils down to the minimization of an objective function comprising the negative log-likelihood and the negative log-prior terms,
\begin{equation}
    \label{eq:priors}
     \hat{\mathbf{x}} = \mathrm{arg}\min_{\mathbf{x}} \; -\log P_{Y|X} ( \mathbf{y} | \mathbf{x} ) -\log P_X (\mathbf{x}).
\end{equation}
In the case of additive white Gaussian measurement noise, the first term becomes the Euclidean norm, $\| \mathcal{F}(\mathbf{x})-\mathbf{y}\|^2$.
Famous examples of prior terms include total variation (TV)~\cite{rudin1992nonlinear} or sparsity with respect to some dictionary~\cite{Lustig2007Sparse}, which are regularly employed in medical imaging.

\subsection{Supervised learning for inverse problems}
As illustrated in Fig. \ref{fig:framework} (top), given an aligned set of samples of latent signals and their corresponding measurements $\{(\mathbf{x}_i,\mathbf{y}_i)\}_{i=1}^{N}$, supervised learning methods aim at estimating the inverse operator that maps the measurements to the corresponding latent signals. We denote by $\mathcal{I}_\theta$ the inverse operator that should invert the action of the forward model. The set of parameters $\theta$ denotes the trainable degrees of freedom -- in our case, the weights of the reconstruction neural network. The training is carried out by minimizing the empirical loss 
\begin{equation}
    \label{eq:supervised_loss}
    \min_{\theta} \; \sum_{i=1}^{N} \mathcal{L}\left(\mathcal{I}_\theta(\mathbf{y}_i), \mathbf{x}_i \right)
\end{equation}
where $\mathcal{L}$ measures the discrepancy between the estimated latent signal $\mathcal{I}_\theta(\mathbf{y}_i)$ and the groundtruth $\mathbf{x}_i$. Typical choices include the Euclidean and the $L_1$ distances. In practice, for image restoration tasks, the operator $\mathcal{I}_\theta$ is modeled as a convolutional neural network and the objective \eqref{eq:supervised_loss} is minimized using stochastic gradient-based solvers. Once the optimal set of parameters $\theta^*$ has been learned on the training set, the inverse operator $\mathcal{I}_{\theta^*}$ is applied to solve the inverse problem with previously unseen inputs. 
\subsection{Self-supervised learning}
The focus of this study is the cases where an aligned set of measurements and latent signals is not available or challenging to obtain at the required size (typical supervised training scenarios demand a very large $N$). In the extreme of such cases, one has access to just one sample of measurements $\mathbf{y}$ and the forward operator $\mathcal{F}$. This exact problem has been traditionally tackled by the prior-based methods that we discussed in the Section \ref{sec:prior}. However, a drawback of many prior-based approaches is the need to induce explicit priors on the image instead of learning image- and task-specific priors. On the other hand, in \cite{ulyanov2018deep} and \cite{shocher2018zero}, the authors demonstrated that CNNs by their very own structure can induce a good prior on natural images. The key idea of the proposed self-supervised approach is to find a latent image that is the output of the parametrized inverse operator $\mathcal{I}_\theta$ that best explains the given measurement. Following the ideas in \cite{ulyanov2018deep} and extending them to a general inverse problem setting, our approach can be formalized as the following optimization problem
\begin{equation}
    \label{eq:self_loss}
    \min_{\theta} \; \mathcal{L}\left(\mathcal{F}(\mathcal{I}_\theta({\mathbf{y}})\right), \mathbf{y}).
\end{equation}
Note that the loss function now operates on the measurement space.

Solving the above optimization problem yields $\hat{\bb{x}} = \mathcal{I}_\theta (\mathbf{y})$, i.e., the latent signal at the intermediate stage, as illustrated in Fig. \ref{fig:framework} (bottom). Intuitively, we are searching for an image $\mathbf{\hat{x}}$ that is parametrized by $\{ \theta, \mathbf{y} \}$ that best explains the measurement $\mathbf{y}$ we have in hand. This approach is referred to as \emph{self-supervised} because the measurements themselves provide the supervision to solve the inverse problem by exploiting the prior induced by the CNN. 

\section{Problem setup}
We demonstrate the applicability of the above discussed self-supervised solvers on the task of accelerated MRI reconstruction. In accelerated MR imaging, the field-of-view (FOV) is scanned with a reduced number of measurements that can be achieved by acquiring less data in the $k$-space (Fourier domain) leading to shorter trajectories, which in turn lead to shorter acquisition times. One standard way of designing such acceleration schemes is by acquiring random Cartesian trajectories (that is, directions aligned with the spatial frequency axes) in the $k$-space.

The forward model of accelerated MRI can be therefore faithfully emulated by sub-sampling the fully sampled $k$-space, and it is realistic, in this case, to assume that the forward operator is known with high accuracy. Following the terminology described in Section \ref{sec:methods}, we denote the image derived from fully sampled $k$-space as $\mathbf{x}$ (the latent image), and the image obtained through the sub-sampled $k$-space is denoted by $\mathbf{y}$ (the measurement). The forward model can therefore be formalized as follows:
\begin{equation}
    \label{eq:fastmri_forward}
    \mathbf{y} = \mathcal{F}(\mathbf{x}) = {\mathbf{\Phi}}^{-1}(\mathbf{S} \odot ({\mathbf{\Phi}}\mathbf{x}))
\end{equation}
where $\odot$ denotes element-wise (Hadamard) product, and ${\mathbf{\Phi}}$ and ${\mathbf{\Phi}^\text{-1}}$ denote the forward and the inverse Fourier transforms, respectively. The binary matrix $\mathbf{S}$ denotes the sampling operator that embodies the Cartesian trajectories through which the measurements were obtained; we refer to the rate of decimation induced in $k$-space as the acceleration factor ($AF$). 

\begin{figure}[!t]
    \centering
       \addtolength{\tabcolsep}{-1pt}
    \begin{tabular}{cccc}
    (a) SR $\times$4  &  (b) Dealiasing $\times$4 & (c) SR $\times$8 & (d) Dealiasing $\times$8 \\
        \includegraphics[width=0.24\textwidth]{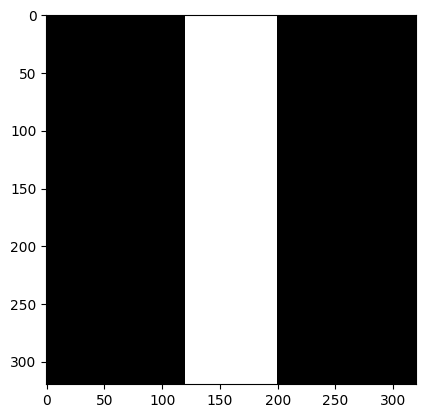}&
    \includegraphics[width=0.24\textwidth]{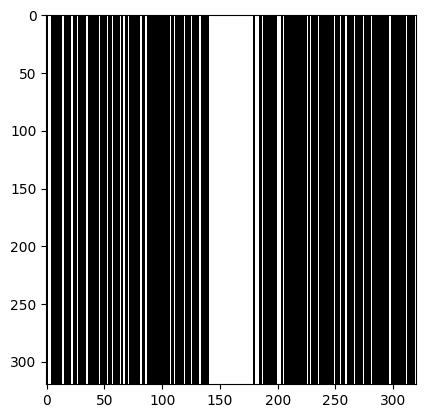}&
        \includegraphics[width=0.24\textwidth]{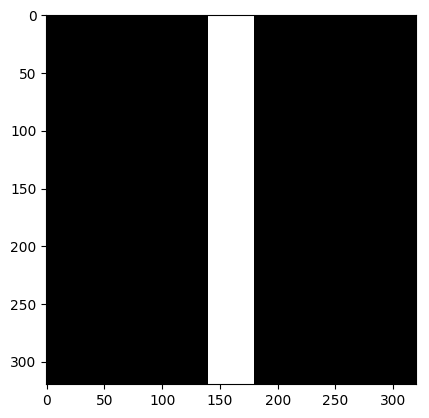}&
    \includegraphics[width=0.24\textwidth]{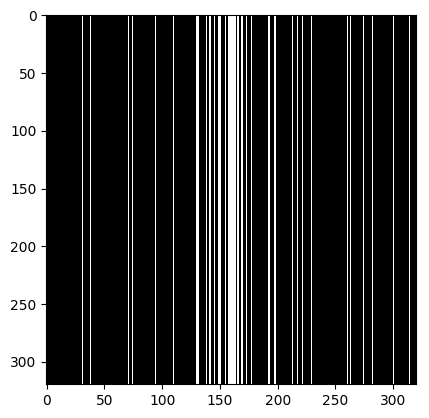}
    \end{tabular}\\
       \addtolength{\tabcolsep}{1pt}
    \caption{{\small Visual comparison of the \textit{k}-space binary masks chosen for different tasks. Tasks ordered from left to right: superresolution $\times 4$, dealiasing $\times 4$, superresolution $\times 8$, and dealiasing $\times 8$  }}
    \label{fig:masks}
\end{figure}

We consider the following two inverse problems:
(i) \textit{Superresolution (SR)}, consisting of reconstructing a sharp image from measurements containing only the central low frequencies obtained by using the mask $\mathbf{S}$ as in Fig. \ref{fig:masks} (a \& c); and 
(ii) \textit{Dealiasing}, in which the obtained mask results in an aliasing artifact due to a coarser sampling in the phase-encoding direction. We use the masks displayed in Fig. \ref{fig:masks}(b \& d). The inverse problem consists of restoring a finer sampling grid in the phase-encoding direction. Throughout the paper, we denote the experiments specifying the task name (one of the two tasks above) and the acceleration factor.  


\paragraph{Loss function.} We use the following loss function:
\begin{align}
    \label{eq:full_loss}
    \mathcal{L}(\mathbf{y},\mathbf{ \hat{y}}) &= \alpha \|\mathbf{y} - \mathbf{\hat{y}}\|_{1} + \beta \| \mathbf{\Phi}\mathbf{y}  - \mathbf{S} \odot \mathbf{\Phi}(\mathbf{\hat{x}})\|_{1} + \gamma \| \mathcal{I}_\theta(\mathbf{y}) - \mathcal{I}_\theta(\mathbf{\hat{y}}) \|_1
\end{align}
comprising three terms. The first term essentially treats the task as a \textit{superresolution} problem, enforcing a penalty on the discrepancy between the reconstructed measurement $\mathbf{\hat{y}}$ and the given measurement $\mathbf{y}$. The second term treats the task as an \textit{inpainting} problem in the \textit{k}-space,  penalizing the discrepancy between the masked Fourier transform of the reconstructed latent image and $k$-space representation of the measurements.
The last term enforces cycle consistency on the reconstructed measurement image passing it through the inverse operator ($\mathcal{I}_\theta(\mathbf{\hat{y}})=\mathbf{\tilde{x}}$) and making it consistent with that of the original measurements ($\mathcal{I}_\theta(\mathbf{y}) = \mathbf{\hat{x}}$). This constraint is similar to the cycle-consistency loss used in \cite{CycleGAN2017} for image style transfer.
In all the three cases, the $L_1$ norm measures the discrepancy; the relative importance of each term is governed by the parameters $\alpha,\, \beta$, and $\gamma$.

\section{Experiments and discussion}
\subsubsection{Compared algorithms.} The proposed SSL scheme was compared to the following two baselines:
(i) \textit{Total variation}: Similarly to \cite{zbontar2018fastmri}, we compare our results to an off-the-shelf accelerated MR reconstruction method with a total variation (TV) regularizer. 
We used the BART \cite{ESPIRiT2014} toolkit for calculating the TV-based MR image reconstruction. The regularization weight was set to $0.01$, and it was run for $200$ iterations per slice.  
(ii) \textit{Supervised learning}: Since the results of a supervised restoration model would be considered as an upper bound on SSL's performance, we trained a U-Net model \cite{ronneberger2015u} on a dataset of aligned reduced measurements and full measurements MRI, and compared our results on samples that were excluded from the training set.      
\subsubsection{Data.}
The data used in the preparation of this article were obtained from the NYU fastMRI Initiative database (\url{fastmri.med.nyu.edu}) \cite{zbontar2018fastmri}. 
As such, NYU fastMRI investigators provided data but did not participate in analysis or writing of this report. 
We have used the fastMRI training set for our experiments and generated two separated sets out of it: one containing $973$ volumes ($34700$ slices) for training and validation, and one containing $8$ volumes ($48$ slices) for testing. Only samples from the test set were used for evaluating all methods: both SSL and the comparison baselines.


\begin{figure}[!htb]
   \centering
   \addtolength{\tabcolsep}{-1pt}
\begin{tabular}{cccccc}
Groundtruth & Corrupted & TV \cite{ESPIRiT2014} & SSL (ours) & Supervised \cite{zbontar2018fastmri} &\\
\includegraphics[width=0.18\textwidth]{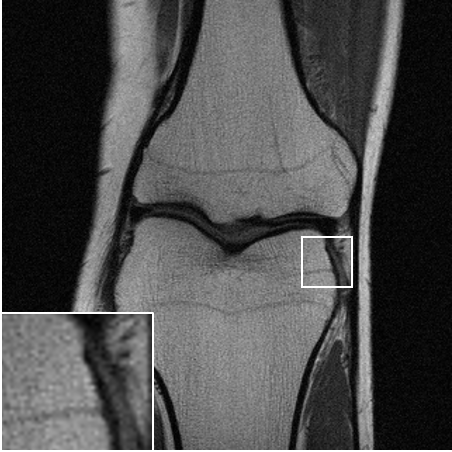}&
\includegraphics[width=0.18\textwidth]{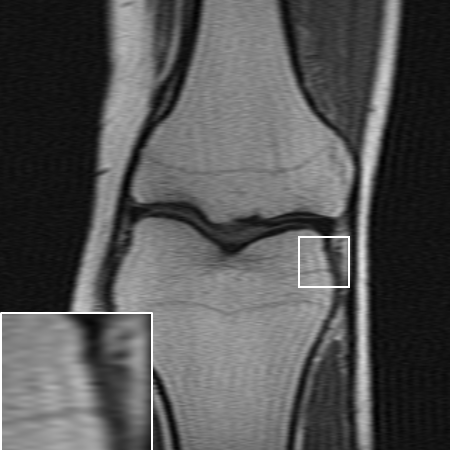}&
\includegraphics[width=0.18\textwidth]{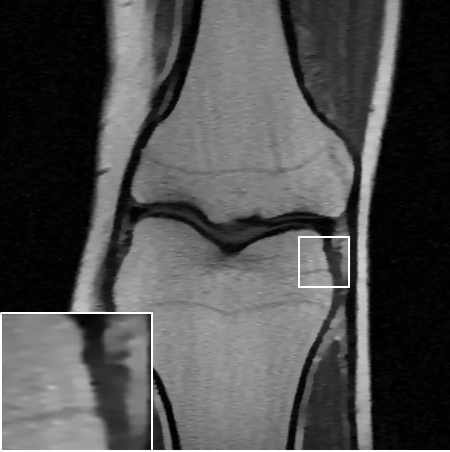}&
\includegraphics[width=0.18\textwidth]{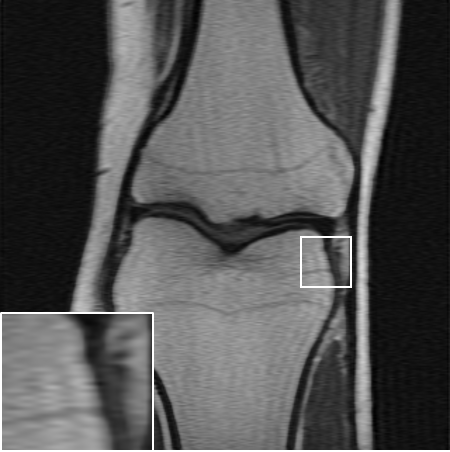}&
\includegraphics[width=0.18\textwidth]{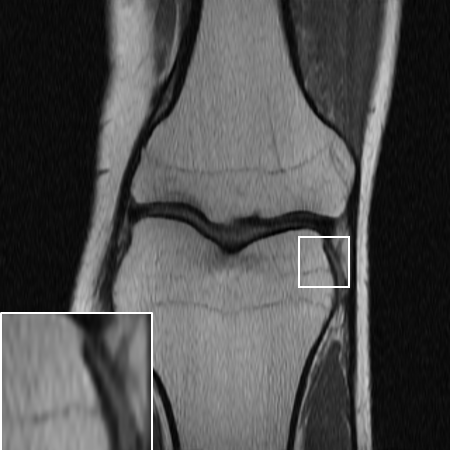}&
\rotatebox{90}{  \; \; \textit{SR}  $\times 4$}
\\
& 25.52dB, 0.683 &  27.14dB, 0.667 & 29.44dB, 0.747 & 30.51dB, 0.769 \\

\includegraphics[width=0.18\textwidth]{results/gt.png}&
\includegraphics[width=0.18\textwidth]{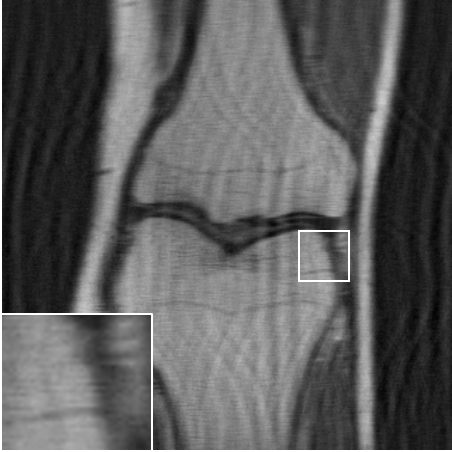}&
\includegraphics[width=0.18\textwidth]{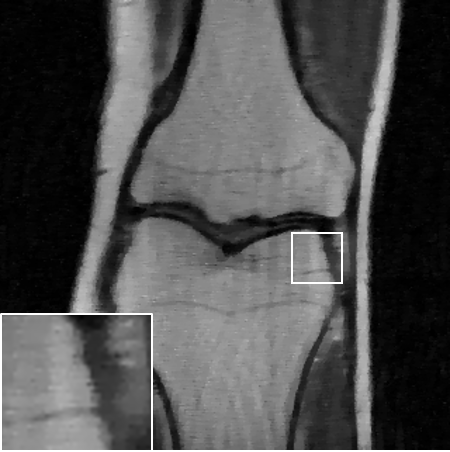}&
\includegraphics[width=0.18\textwidth]{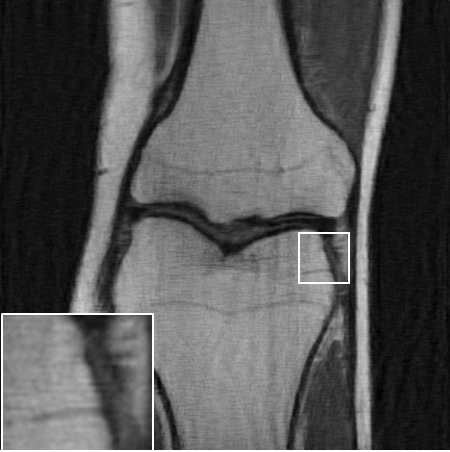}&
\includegraphics[width=0.18\textwidth]{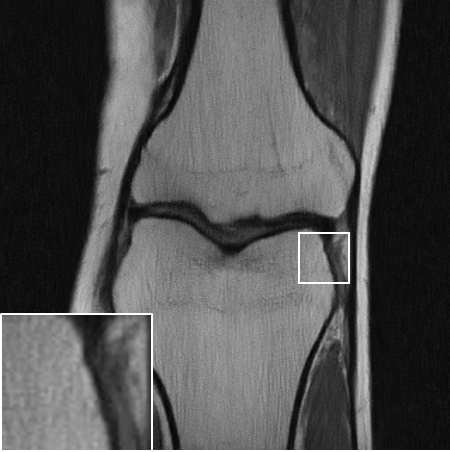}&
\rotatebox{90}{\, Dealiasing $\times 4$}
\\
& 21.18dB, 0.577 &  26.17dB, 0.626 & 29.13dB, 0.719 & 30.51dB, 0.786 \\

\includegraphics[width=0.18\textwidth]{results/gt.png}&
\includegraphics[width=0.18\textwidth]{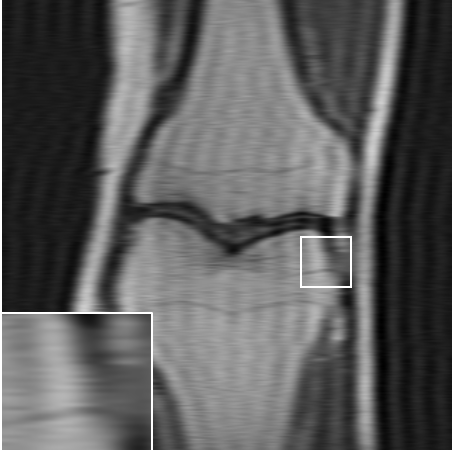}&
\includegraphics[width=0.18\textwidth]{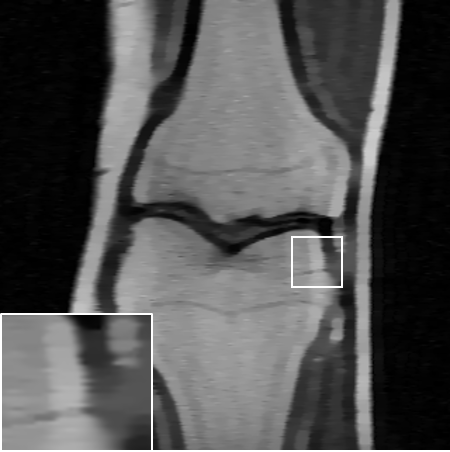}&
\includegraphics[width=0.18\textwidth,trim={1mm 0 0 0},clip]{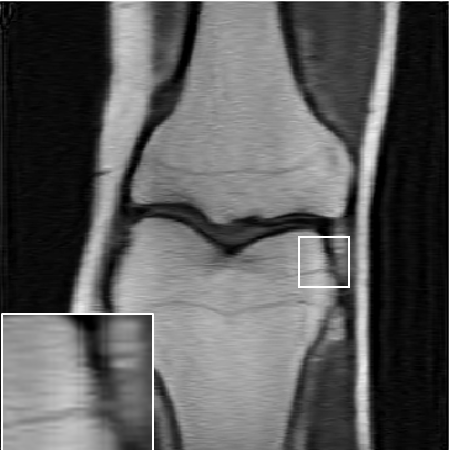}&
\includegraphics[width=0.18\textwidth]{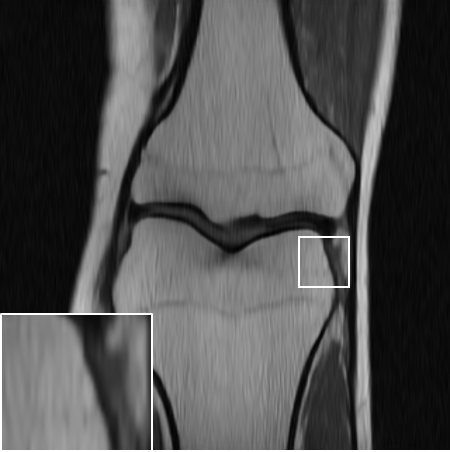}&
\rotatebox{90}{  \; \; \textit{SR} $\times 8$ }
\\
& 20.75dB, 0.552 &  23.43dB, 0.541 & 26.39dB, 0.614 & 29.27dB, 0.70 \\

\includegraphics[width=0.18\textwidth]{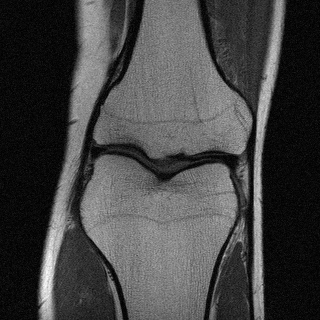}&
\includegraphics[width=0.18\textwidth]{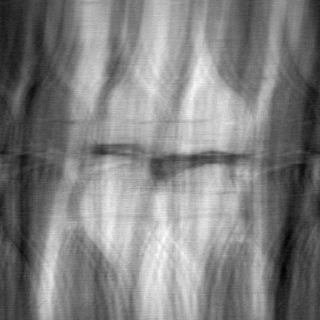}&
\includegraphics[width=0.18\textwidth]{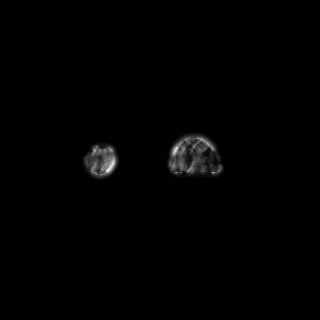}&
\includegraphics[width=0.18\textwidth]{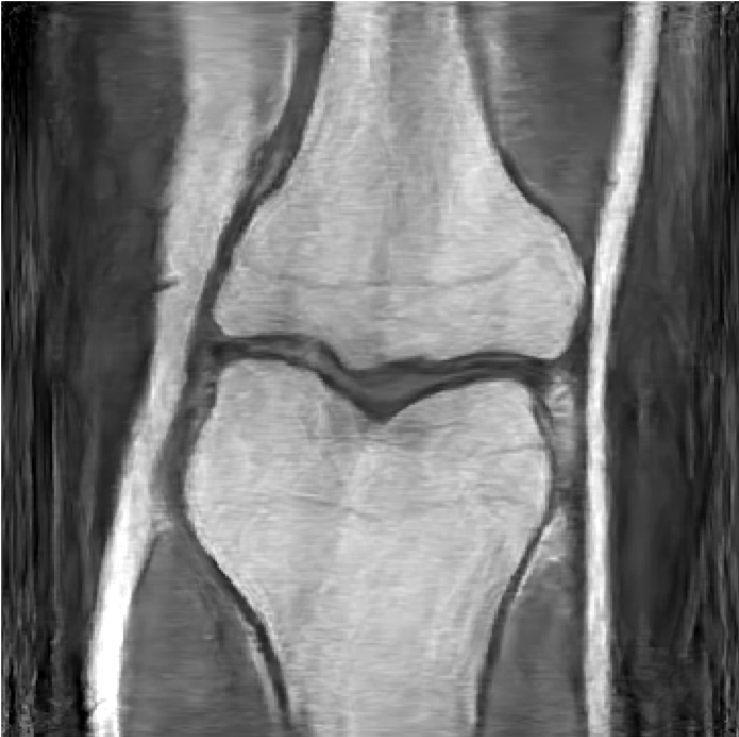}&
\includegraphics[width=0.18\textwidth]{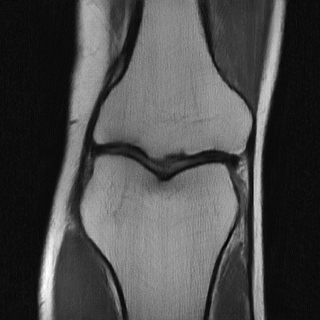}&
\rotatebox{90}{\, Dealiasing $\times 8$}\\
& 11.52dB, 0.348 &  N/A, N/A & 23.27dB, 0.501 & 28.01dB, 0.686 \\

\end{tabular}
\addtolength{\tabcolsep}{1pt}
    \caption{{\small Comparison of the proposed self-supervised approach (SSL) to ESPIRiT (TV)\cite{ESPIRiT2014} and supervised trained network \cite{zbontar2018fastmri} on different tasks. From top to bottom the panels depict the tasks: \textit{SR} $\times 4$, dealiasing $\times 4$,  \textit{SR} $\times 8$ and dealiasing $\times 8$. From left to right, the columns depict the groundtruth, corrupted (with respective masks), TV-restored \cite{ESPIRiT2014}, SSL-restored(\textit{ours}), and supervised model restored images, along with their corresponding (PSNR, SSIM) metrics mentioned below. }} 
    \label{fig:vis_results} 
\end{figure}
\subsubsection{Settings.}
We  performed our experiments with various types of inputs to the inverse operator $\mathcal{I}_\theta$: the measurements themselves $\mathbf{y}$, a gradient-like image which we refer to as "meshgrid" input ($\mathbf{z}$) -- similar to what has been used in \cite{ulyanov2018deep}, and a combination of the measurements and the meshgrid input -- stacked as two channels, i.e., $[\mathbf{y} \;\;\mathbf{z}]$. Since cycle-consistency is not valid in the case of a meshgrid (only) input, $\gamma$ has been set to zero for these experiments. 
The inverse operator $\mathcal{I}_\theta$ was chosen to be the U-net architecture for all our experiments~\cite{ronneberger2015u}. We used the Adam optimizer \cite{Adam2014} as the update method with learning rate of $10^{-4}$.  
\subsubsection{Results.}
Table \ref{tab:quant_results} summarizes the results for the various experimental settings, comparing the performance of the above-mentioned benchmarks to ours. When compared to TV~\cite{ESPIRiT2014}, our method achives an improvement of $\sim 2$dB-$3$dB in the peak-signal-to-noise ratio (PSNR). Similar results are observed for the structural-similarity measure (SSIM, about $0.05-0.08$ points improvement). As expected, the supervised model outperforms the proposed SSL method. However, it seems that at least for the lower distortion rates, this gap is surprisingly small. A visual inspection of the results over one slice is provided in Fig. \ref{fig:vis_results}.  As can be observed in the zoomed-in region, our method manages to restore finer details better than the TV-based method, and even approaches the restoration levels of the supervised model in the lower distortion rates. As evident both quantitatively and visually, the TV-based method completely fails on the $\times 8$ dealiasing task, whereas our SSL method seems to significantly alleviate the reconstruction artifacts.   
\begin{table}[t]
\centering\setlength\extrarowheight{2pt}
\caption{{\small Quantitative evaluation of the proposed method on 48 slices from 8 volumes. The volumes were chosen randomly from the validation set of the fastMRI dataset \cite{zbontar2018fastmri}.}}
\begin{tabular}{@{\extracolsep{4pt}}llc*{2}{c@{\enspace}c@{\enspace}c}}
\hline
 Task & Metrics & Corrupted & TV & SSL (ours) & Supervised  \\
   \hline
   \multirow{2}{*}{Superresolution $\times 4$}
& PSNR & 25.25 & 25.61 & 28.08 & 28.79 \\
& SSIM & 0.683 & 0.627 & 0.691  & 0.701 \\
\hline
\vspace{1pt}  
    \multirow{2}{*}{Dealiasing $\times 4$}
  & PSNR &  21.88 & 25.26 & 27.56 & 28.67 \\
  & SSIM &  0.587 & 0.579 & 0.66  & 0.7056 \\
\hline
\vspace{1pt}  
    \multirow{2}{*}{Superresolution $\times 8$}
  & PSNR & 22.16 & 23.29 & 25.61 & 27.25 \\
  & SSIM & 0.5224 & 0.5017 & 0.5541  & 0.6057 \\
\hline
\vspace{1pt}
  \multirow{2}{*}{Dealiasing $\times 8$}
  & PSNR & 13.74 & N/A & 22.86 & 26.72 \\
  & SSIM & 0.40 & N/A & 0.47  & 0.604 \\
 \hline
 \label{tab:quant_results}
\end{tabular}
\end{table}
\subsubsection{Discussion.} 
From the practical perspective, we observed that for lower decimation rates ($\times 4$) using the input as the measurements $\mathbf{y}$ (or) $[\mathbf{y} \; \; \mathbf{z}]$ yielded the best performance. At higher decimation rates ($\times 8$), we observed that using $\mathbf{z}$ as the input performs better than using the measurements. \cite{ulyanov2018deep} observed a similar behavior: different restoration tasks required different inputs. This implies that the input is part of the induced prior, and specifically in our case, the restoration tasks involving higher decimation rates require a smooth input that is distinct from the distorted measurements.

Upon performing a hyper-parameter search for ($\alpha, \beta$, $\gamma$), we observed that, irrespective of the input, enforcing a higher weight on the \textit{k}-space loss is crucial relative to the spatial loss. In all $\times 4$ experiments we used $(\alpha, \beta, \gamma) = (1.0, 8.0, 10^{-5})$, while for the $\times 8$ experiments we set them to $(0.0, 7.0, 0.0)$.
\section{Conclusion and future work}
We proposed a new learning framework for solving inverse problems in the absence of aligned data. As a proof-of-concept, we demonstrated the applicability of the proposed framework to the use case of accelerated MRI reconstruction, where our approach outperforms standard off-the-shelf solvers by a significant margin. We believe this framework leads to many interesting future directions and can become a tool in solving a new range of inverse problems in the limited/no aligned data regime where the supervised methods are not applicable. One could devise better loss functions that can be imposed in the measurements domain. This framework could be extended to a semi-supervised scenario as well and used for analysing the importance of external (supervised) data.

\section*{Acknowledgement}
This research was supported by ERC StG RAPID.
%
%
%
\bibliographystyle{splncs04}
\bibliography{self_supervisedMain}

\end{document}